\documentclass[journal]{IEEEtran}

\pdfoutput=1
\usepackage{xcolor,soul,framed} 

\colorlet{shadecolor}{yellow}
\usepackage[pdftex]{graphicx}
\graphicspath{{../pdf/}{../jpeg/}}
\DeclareGraphicsExtensions{.pdf,.jpeg,.png}

\usepackage{bm}

\usepackage[cmex10]{amsmath}
\usepackage{array}
\usepackage{mdwmath}
\usepackage{mdwtab}
\usepackage{eqparbox}
\usepackage{url}
\usepackage{makecell}
\usepackage{caption}
\usepackage{dutchcal}
\usepackage{upgreek}
\usepackage{color}

\begin{document}
\bstctlcite{IEEEexample:BSTcontrol}
    \title{Enhancement of Direct LEO Satellite-to-Smartphone Communications by Distributed Beamforming}
  \author{Zhuoao Xu,~\IEEEmembership{Student Member,~IEEE,}
      Gaojie~Chen,~\IEEEmembership{Senior~Member,~IEEE,}\\
      Ryan~Fernandez,
      Yue~Gao,~\IEEEmembership{Senior~Member,~IEEE,}
      and~Rahim~Tafazolli,~\IEEEmembership{Senior~Member,~IEEE}

\thanks{Z. Xu, G. Chen, R. Fernandez, Y. Gao and R.Tafazolli are with the Institute for Communication Systems (ICS), 5GIC \& 6GIC, University of Surrey, Guildford, Surrey, GU2 7XH, U.K. (email:\{z.xu, gaojie.chen, r.fernandez, r.tafazolli\}@surrey.ac.uk; yue.gao@ieee.org).
(Corresponding author: Yue Gao) 

} }  


\maketitle

\begin{abstract}
The low earth orbit (LEO) satellite network is undergoing rapid development with the maturing of satellite communications and rocket launch technologies, and the demand for a global coverage network. However, current satellite communication networks are constrained by limited transmitting signal power, resulting in the use of large-size and energy-consuming ground terminals to provide additional gain. This paper proposes a novel technology called distributed beamforming to address such challenges and support direct communications from LEO satellites to smartphones. The proposed distributed beamforming technique is based on the superposition of electromagnetic (EM) waves and aims to enhance the received signal strength. Furthermore, we utilize EM wave superposition to increase the link budget and provide the coverage pattern formed by the distributed antenna array, which will be affected by the array structure and the transmitter parameters. In addition, the impact of Doppler frequency shift and time misalignment on the performance of distributed beamforming is investigated. Numerical results show that the enhancement of the received power depends on the angle formed by those radiated beams and can be up to the square of the number of beams; namely, a maximum enhancement of 6 dB could be obtained by using two satellites and a maximum of 12 dB increase through four satellites, which provide a clear guideline for the design of distributed beamforming for future satellite communications.

\end{abstract}

\begin{IEEEkeywords}
Distributed beamforming, LEO mega constellation, LEO satellite-to-smartphone communications, Distributed antenna array.
\end{IEEEkeywords}

%
\IEEEpeerreviewmaketitle


\section{Introduction}

\IEEEPARstart{S}{atellite} communication plays a vital role in wireless communication because of its appealing features, such as large coverage area, high stability, and high reliability. According to the orbital altitude, satellites can be classified into geostationary (GEO) satellites with a constant altitude of 35786 km, medium earth orbit (MEO) satellites with an altitude between 2000 and 35786 km, and low earth orbit (LEO) satellites with an altitude of ranging from about 500 km to 2000 km. In recent years, research on LEO satellites has been increasing, including integration with terrestrial mobile communication \cite{Integration, Terrestrial}, and collaboration with GEO \cite{GEO1, GEO2}. Cost reduction is the most critical factor driving the development of the LEO mega constellation. Rocket reuse and multiple satellites launched via a rocket reduce the launch cost immensely. Another essential reason is the need for networks in remote areas. Compared with MEO and GEO communication networks, LEO has priorities like lower latency and smaller path loss  \cite{6Gtrend}.

\begin{table}[t]
 \centering
 \caption{Link Budget.} 
     \begin{tabular}{c|c}

    \hline
     \hline
       \textbf{ Parameters}  &  \textbf{ Values}  \\
        \hline
         \hline
         
       Distance  &  600 km \\
    \hline
       Operating frequency  &  3.5 GHz \\
    \hline
       Free space path loss (FSPL)  &  158.9 dB \\
    \hline
       Effective isotropic radiated power (EIRP)  &  36.7 dBW \\
    \hline
       Transmit antenna gain  &  37.1 dBi \\
    \hline
       Receive antenna gain  &  0 dBi \\
    \hline
       Atmospheric losses and Rain attenuation  &  5 dB \\
    \hline
       Losses from transmitter  &  2 dB \\
    \hline
       Losses from receiver  &  2 dB \\
    \hline
       Received power calculated & -101.2 dBm \\
    \hline
     \end{tabular}
 \end{table}
 
LEO satellites and the space Internet attract much attention in both industry and academia. Among the LEO mega-constellations, including One Web and Amazon's Kuiper, Starlink from SpaceX is taking the lead in building the space network \cite{4Constellation}. Even though its communication system still needs a large device to receive signals from satellites and transmit them to user equipment (UE). The receiving device is a dish antenna whose latest version is around 0.5 meters long and 0.3 meters wide. Some other companies, such as LYNK Global and Applied Satellite Technology (AST), focus more on satellite-to-mobile communications technology than constellation building. LYNK successfully realized two-way direct communication between an LEO satellite and a standard mobile phone \cite{LYNK}, but only short message service (SMS) can currently be provided due to the limited network quality. AST is trying to deploy massive arrays of antennas in space to provide broadband Internet.

Suppose that a standard 5G smartphone operating at 3.5 GHz is used to substitute the original satellite receiving apparatus to receive signals from satellites directly. The reason for selecting this frequency is its capability to strike an optimal balance between coverage and capacity. This decision takes into account the higher path loss experienced in Ku and Ka bands, as well as the lower data rate observed in the S-band. As calculated in Table I, the received power can reach up to -101.2 dBm. However, such a signal strength is too weak for many applications. From 3GPP TS 38.101-1 \cite{3GPP}, the minimum reference sensitivity for operating band n78 is -96.5 dBm. To enhance the received power, one can opt to either increase the transmission power or gain of a single satellite or utilize multiple satellites in collaboration. Considering the cost and flexibility, multi-satellite cooperation should be a better solution. There have been various studies targeting satellite multiple-input multiple-output (MIMO) systems aimed at providing diversity gain or multiplexing gain to enhance system-level performance. However, these studies often overlook the challenge posed by weak received signals, which makes direct reception by smartphones difficult. To address this limitation, we propose a novel multi-satellites based distributed beamforming that significantly enhances received power. The distributed beamforming technique relies on the superposition of electromagnetic (EM) waves, allowing for a substantial increase in received signal strength. This enables effective transmission while preserving the portability of individual satellites.

In summary, the proposed distributed beamforming, designed for the LEO mega-constellation, offers a novel solution for standard smartphones to access LEO satellite networks. It focuses on improving the received signal strength through distributed beamforming, considering both propagation and antenna aspects. The main contributions of this paper can be summarized as follows:

\begin{itemize}
\item A novel distributed beamforming technique based on the LEO mega-constellation is proposed for the first time, addressing the challenge of insufficient received power and enabling direct communication from LEO satellites to the UE on Earth.

\item The study investigates the enhanced received signal strength achieved through distributed beamforming in various scenarios, utilizing the theory of EM wave propagation and EM field superposition. Additionally, the analysis considers the impact of offsets, such as Doppler frequency shift and time misalignment, on the performance of distributed beamforming.

\item Numerical simulation results demonstrate that the combination of distributed beamforming with LEO mega-constellations can achieve a maximum of $N^2$-fold enhancement of received power with N satellites, surpassing the improvement achieved by traditional satellite MIMO techniques, which provides only an $N$-fold increase. For instance, through distributed beamforming, a configuration with two satellites moving in parallel can achieve a maximum of 6 dB received power enhancement, while four satellites moving vertically could obtain a 9 dB increase, and four satellites in parallel would experience a remarkable 12 dB improvement.
\end{itemize}  

The rest of this article is organized as follows. Section II presents related work on satellite MIMO, distributed beamforming, and synchronizations. Section III depicts the system models, including the LEO satellite networks, phased array, distributed array, and Doppler effect models. Then in Section IV, the enhancement of received signal strength by distributed beamforming is comprehensively analyzed in different scenarios such as parallel orbits, perpendicular orbits, and intersecting orbits. Section V gives the numerical simulation results and coverage patterns in different cases. Doppler effect and the effect of time misalignment are also discussed. Finally, we conclude this work in Section VI.

\section{Related works}

\subsection{Satellite MIMO}
In \cite{LMSMIMO}, the authors studied a land mobile satellite (LMS) MIMO model where two satellites communicate simultaneously with the same mobile UE. They adopted space-time block coding for signal transmission between satellites and the UE, resulting in better bit error rate performance than single satellite communications. A novel cell-free massive multiple-input multiple-output (CF-mMIMO) architecture was introduced for future ultra-dense LEO satellite networks \cite{CFMIMO}. The authors discussed network design, power allocation, and handover management processes, aiming to increase average service time and spectral efficiency. However, they did not consider the concrete realization of beamforming in their work. The authors also proposed an AI-based implementation for joint optimization of power allocation and handover management processes based on distributed massive MIMO (DM-MIMO), taking into account real-time operation and the dynamic satellite network environment \cite{DM-MIMO}. Their focus was on framework optimization and reducing handover rates. In \cite{OCMIMO}, an approach was presented to achieve maximum MIMO spectral efficiency for satellite-to-ground communications by optimizing the displacement of antennas on both sides. The analysis in \cite{UHF} explored the potential application of MIMO spatial multiplexing technology in UHF satellite communication systems. They assumed a powerful receiver to ensure smooth signal reception. A novel passive measurement technique was presented in \cite{SATCOM-MIMO} to validate the orthogonality of the MIMO satellite communication channel matrix. This validation is beneficial for promoting the application of MIMO in satellite communication. To improve beamspace MIMO for satellite swarms, the authors of \cite{BeamspaceMIMO} proposed a distributed linear precoding scheme combined with a ground station equalizer to reduce dependence on perfect channel state information (CSI) and strong coordination between satellites. However, the satellite MIMO-related techniques used in the above works mainly focus on spatial diversity, spatial multiplexing, and a few users, which leads to limited application scenarios and a small receiving gain. For example, the gain can be improved by up to $N$ times when $N$ satellites work coherently, which may not be enough for high-speed transmission.

\subsection{Distributed Transmit Beamforming}
Beamforming is an effective method for achieving high transmitting gain as it allows coherent superposition of energy in the desired direction. The feasibility of adopting distributed beamforming in wireless networks was discussed in \cite{FeasibilityDB}. In this approach, a cluster of distributed sources coherently transmits a standard message signal to the same target. Despite moderate phase errors, a large fraction of the beamforming gain can still be achieved. The concept of distributed transmit beamforming was further analyzed in \cite{CBF} and \cite{DTB}. In \cite{CBF}, a detailed theoretical analysis was presented, while \cite{DTB} specifically introduced techniques for carrier phase synchronization. Separate investigations of distributed phased arrays with and without destination feedback were conducted in \cite{DBwithFB} and \cite{DBwithoutFB}, respectively. In \cite{Swarm}, the authors proposed a "Satellite Swarm" structure, composed of small and light satellites (e.g., CubeSats) that directly connect with smartphones. The primary contribution of this work lies in the geometrical design of the swarm array, which effectively mitigates the grating lobes even with an element spacing of one meter (equivalent to 6.67 wavelengths at 2 GHz). An antenna technology called "Formation-of-Arrays (FOA)" was introduced in \cite{FOA}, aiming to obtain sufficiently high throughput. However, this technique requires satellites to be relatively close to each other. For instance, in the GEO case, the element spacing should be 4.5 wavelengths, whereas in the LEO case, it should be only 0.6 wavelengths. It should be emphasized that the distributed beamforming techniques discussed in the aforementioned studies resemble conventional arrays, wherein the spacing between transmitters or receivers is on the same scale as the wavelength. In contrast, in LEO satellite networks, satellites, acting as radiation sources, are usually positioned a few kilometers or more apart.

\subsection{Synchronization}
The synchronization of signals among cooperating satellites plays a vital role in the proper operation of the system, enabling a coherent combination of received signals at the receiver. In a previous study \cite{Clock1}, one-way wireless clock transfer was employed to align clocks among transmitters by modulating the clock signal onto a carrier frequency, although it relied on high-accuracy microwave ranging. Various synchronization methods for time, frequency, and phase were described in \cite{DTB}, encompassing both closed-loop and open-loop approaches. Leveraging the known locations and motion of nodes, radios can calculate the phases required for beamforming \cite{Phase}. A novel approach catering to high-mobility platforms, such as unmanned aerial vehicles (UAVs), was proposed in \cite{Guided} for synchronization without the need for destination feedback. However, this method assumes that the distributed radios are in close proximity to each other.

Furthermore, LEO satellites experience a more severe Doppler effect compared to other satellites, primarily due to their lower orbits and higher velocity. The Doppler effect refers to the signal frequency difference between the transmitter and the receiver caused by their relative motion. Its characterization for LEO satellites has been presented in \cite{DooplerLEO}, where the Doppler frequency shift is calculated in the earth-centered fixed (ECF) coordinate frame. In light of the Doppler effect, the authors of \cite{DooplerBeamSize} and \cite{DopplerMIMO} integrated Doppler compensation into their investigations on adaptive beam design and distributed LEO-MIMO satellite communications. Despite efforts to compensate for the frequency offset at one point, it may reappear at another point significantly distant from the reference location. To address this issue, the authors illustrated the application of precoding to an Orthogonal Time and Frequency Space (OTFS) waveform in \cite{OTFS}, aiming to eliminate residual time and frequency offsets. Whereas, the work assumed that there was direct communication between each satellite and the UE. Given the existence of phase, frequency, and time offsets, it is imperative to consider their impacts on the enhancement of received power in the context of the system.

\section{System Model and Problem Formulation}
\subsection{System Model}
\begin{figure}
  \centering
  \includegraphics[width=3in]{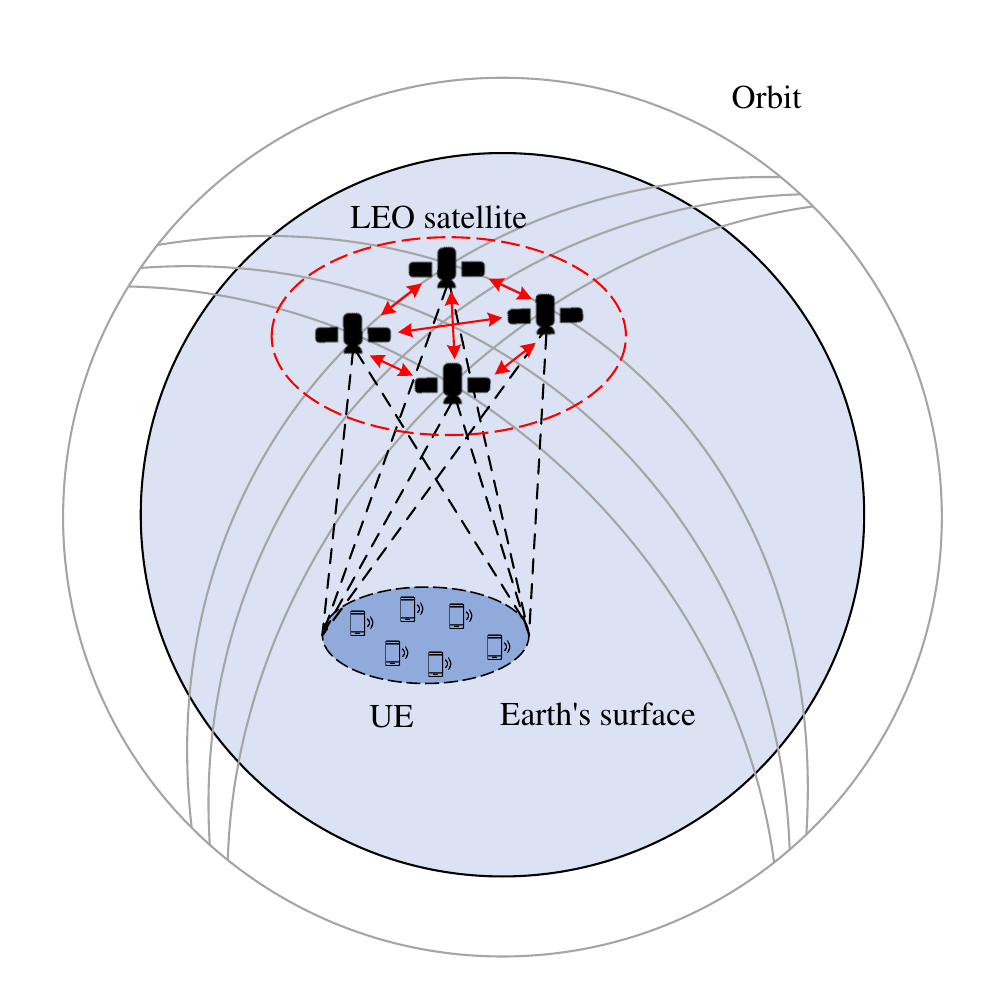}
  \caption{LEO satellite network with distributed beamforming.}\label{Network}
\end{figure}
Fig.~\ref{Network} illustrates the LEO satellite network with distributed beamforming where the outermost circles and curves indicate the satellite orbits and the sphere surrounded by the orbits indicates the Earth. Satellites in the red dashed circle form a distributed array and propagate their EM waves toward the common coverage area. Soon afterward, some satellites move out of the circle, and a few others move into it, forming a new distributed array. The network is established as a single-carrier broadcasting network, wherein a distributed array of satellites transmits identical information to multiple users. This strategic approach ensures the constructive combination of superimposed signals, ultimately leading to a reception level that is easily discernible and comprehensible by the UE. Typically, UE at the earth's surface has one standard receiving antenna, and only the line-of-sight channel model is considered. Distributed arrays do not have the same inherent properties as traditional arrays, such as time alignment and frequency synchronization. Therefore, to achieve good performance through distributed beamforming, time alignment, frequency synchronization, and phase calibration should be satisfied as priorities. Inter-satellite links (ISLs) are adopted to exchange data and ensure synchronization between satellites \cite{ISL1, ISL2, ISL3, ISL4}, represented by those lines with two-way arrows. According to the inclination of the satellite orbit, the orbit can be classified as inclined or polar orbit. Due to the circular orbit, the satellite ascends on one side of the orbit and descends on the other. As a result, ascending satellites and descending satellites could get very close. Furthermore, the mega-constellation is comprised of multiple shells which have different altitudes. Satellites flying in various shells also make the smaller spacing possible.

\subsection{Phased Array Model}
\begin{figure}
  \centering
  \includegraphics[width=3.3in]{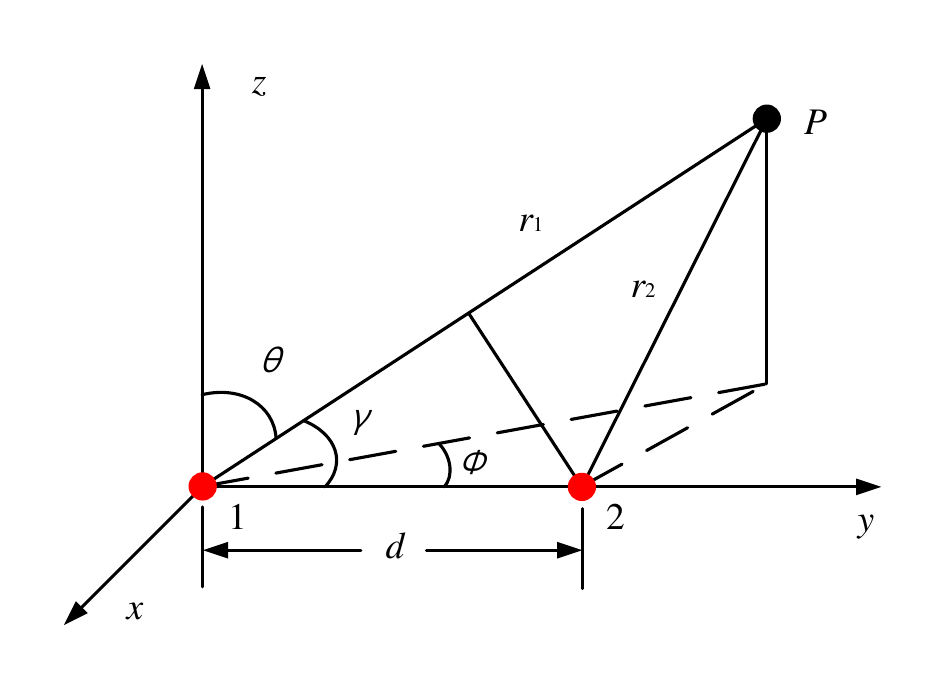}
  \caption{Traditional phased array model.}\label{TA}
\end{figure}
The traditional phased array antenna on the satellite can steer the beam propagation direction as required. Its model is shown in Fig.~\ref{TA}. Source one is located at the original point, and source two is located at the $x$ axis with a distance of $d$ away from source one. $r_{1}, r_{2}$ respectively denotes the distance from source one and source two to the target point $P$. $\theta, \phi, \varphi$ separately represents the elevation angle, the azimuth angle, and the angle between the main radiation direction and $x$ axis. Supposing that the two point sources have the same current amplitude, $I_{m1} = I_{m2}$, but have different phases, $I_{1} = I_{2}e^{\xi}$. Then, the electric (E) field expressions at the target point can be written as
\begin{equation}\label{Traditional E1}
E_{1} = E_{m} f_{e}(\theta,\phi) \frac{e^{-jk r_{1}}}{r_{1}}
\end{equation}
\begin{equation}\label{Traditional E2}
E_{2} = E_{m} f_{e}(\theta,\phi) e^{\xi} \frac{e^{-jk r_{2}}}{r_{2}},
\end{equation}
where $E_{m}$ is the amplitude, $f_{e}$ represents the field pattern of an individual element, and $k$ means the wave number. Thus, the total E field is
\begin{equation}\label{Traditional Etotal}
E_{total} = E_{1} + E_{2} = E_{m} f_{e}(\theta,\phi) \left[\frac{e^{-jk r_{1}}}{r_{1}} + e^{\xi} \frac{e^{-jk r_{2}}}{r_{2}}\right].
\end{equation}

For a traditional array, $d$ is usually less than one wavelength. Considering that the target point is far from radiation sources, two approximations can simplify the calculation. The first one approximates the distance between the radiation source and the target. As $r_{1}, r_{2}$ are much longer than $d$, there is
\begin{equation}
\frac{1}{r_{1}} \approx \frac{1}{r_{2}}.
\end{equation}
The second approximation is that the beam propagates in parallel directions. Therefore, to the target, the distance difference between the two beams can be expressed as
\begin{equation}
r_{1} - r_{2} = d \cos \gamma,
\end{equation}
where
\begin{equation}
\cos \gamma = \sin \theta \cos \phi.
\end{equation}

\subsection{Distributed Array Model}
If LEO satellites are regarded as radiation sources, the characteristics of the distributed array composed of satellites are different from those of the traditional phased array. On the one hand, (4) would not apply to the distributed array as $r_{1}$ and $r_{2}$ could have a large difference. On the other hand, (5) cannot be used because radiation beams are not parallel, causing beam propagation distances to no longer be correlated. The specific orbit altitudes and inclinations cause the spacing between the satellites to change, constantly forming an irregularly distributed array. Besides, the spacing is usually more than a few kilometers, much larger than the operating wavelength.

\begin{figure}
  \centering
  \includegraphics[width=2.8in]{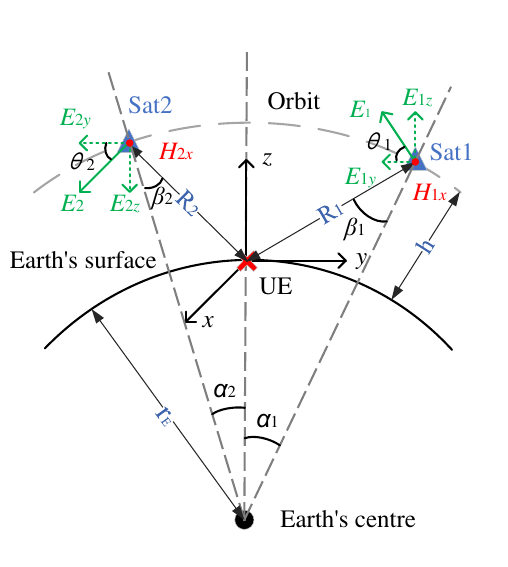}
  \caption{The superposition of EM waves radiated from LEO satellites. The black solid curve represents the earth’s surface, the grey dashed curve above it represents the satellite operating orbit, and the black point below represents the earth’s center.}\label{DA}
\end{figure}

The distributed array model for satellite-to-smartphone is illustrated by Fig.~\ref{DA}, where the simplest case is given with only two satellites moving in the same direction. In this model, we adapt two kinds of coordinates. One is a spherical coordinate system with the earth's center as its origin, and the other is a rectangular coordinate system with the UE as the origin. The connection line between the earth's center and the UE coincides with the z-axis. In the spherical coordinate system, $\alpha_{1}, \alpha_{2}$ represent the polar angles of satellite 1 (Sat1) and satellite 2 (Sat2) respectively. The angle formed by the intersection of Sat1-to-earth's center connection line and Sat1-to-UE is denoted by $\beta_{1}$.
Similarly, the corresponding one for Sat2 is denoted as $\beta_{2}$. The earth's radius $r$ and orbit altitude $h$ are known. If the polar angles $\alpha_{1}$ and $\alpha_{2}$ can be obtained, then all the geometrical parameters in the distributed array model can be attained by calculation as follows.

Through the law of cosines, there is
\begin{equation}
\cos \alpha_{1} = \frac{(r_{E}+h)^{2}+r_{E}^{2}-R_{1}^{2}}{2r_{E}(r_{E}+h)}.
\end{equation}
Rearranging the equation, the distance between Sat1 and UE, $R_{1}$, can be represented as
\begin{equation}
R_{1} = \sqrt{(r_{E}+h)^{2}+r_{E}^{2}-2r_{E}(r_{E}+h)\cos \alpha_{1}}.
\end{equation}
In the same way, $R_{2}$ can also be obtained.
\begin{equation}
R_{2} = \sqrt{(r_{E}+h)^{2}+r_{E}^{2}-2r_{E}(r_{E}+h)\cos \alpha_{2}}.
\end{equation}
Assuming that EM waves transmitted by both satellites have the same initial phase, the phase difference caused by different propagation paths can be written as
\begin{equation}
\varphi_{d} = |R_{1}-R_{2}| \frac{2\pi}{\lambda}.
\end{equation}
Through the law of sines, there is
\begin{equation}
\frac{r_{E}}{\sin \beta_{1}} = \frac{R_{1}}{\sin \alpha_{1}}.
\end{equation}
Rearranging the equation, $\beta_{1}$ can be represented as
\begin{equation}
\beta_{1} = \arcsin(\frac{r_{E} \sin \alpha_{1}}{R_{1}}).
\end{equation}
Then,
\begin{equation}
\theta_{1} = \alpha_{1} + \beta_{1}.
\end{equation}
In the same way, there is
\begin{equation}
\theta_{2} = \alpha_{2} + \beta_{2}.
\end{equation}

EM waves are transverse electromagnetic (TEM) waves when they travel in space, which means the direction of the electric field, magnetic field, and propagation are perpendicular to each other. At first, we know that satellites radiate EM waves to the ground UE. We then assume that the $E$ field is perpendicular to the beam propagation direction to the right, and consequently, the $H$ field is straight into the figure. Then, each satellite's $E$ field and $H$ field can be easily decomposed and composed based on the rectangular coordinate. $\theta_{1}$ and $\theta_{2}$ calculated in (13) and (14) are the decomposition angles. Given these, the components of $E$ fields and magnetic fields could be represented below. Moreover, time domain expression expresses the calculation process more intuitively.
\begin{equation}\label{E}
\begin{cases}
    \bm{E_{1y}} = E_{1m} \cos\theta_{1} \sin(\omega t - k R_{1} + \varphi_{1}) (-\bm{e_{y}}) \\
    \bm{E_{1z}} = E_{1m} \sin\theta_{1} \sin(\omega t - k R_{1} + \varphi_{1}) (\bm{e_{z}}) \\
    \bm{E_{2y}} = E_{2m} \cos\theta_{2} \sin(\omega t - k R_{2} + \varphi_{2}) (-\bm{e_{y}}) \\
    \bm{E_{2z}} = E_{2m} \sin\theta_{2} \sin(\omega t - k R_{2} + \varphi_{2}) (-\bm{e_{z}})
\end{cases}
\end{equation}

\begin{equation}\label{H}
\begin{cases}
    \bm{H_{1x}} = \frac{E_{1m}}{Z_{0}} \sin(\omega t - k R_{1} + \varphi_{1}) (-\bm{e_{x}}) \\
    \bm{H_{2x}} = \frac{E_{2m}}{Z_{0}} \sin(\omega t - k R_{2} + \varphi_{2}) (-\bm{e_{x}}),
\end{cases}
\end{equation}
where $E_{1m}$ and $E_{2m}$ are the $E$ field magnitude at the receiver, $\varphi_{1}$ and $\varphi_{2}$ are the initial phases at the transmitter, $k$ still represents the wave number, and $Z_{0}$ is the wave impedance of free space.

\subsection{Doppler Effect Model}
LEO satellites move in non-geostationary orbit (NGSO), and UE could move at high speed on the ground, causing an inevitable Doppler frequency shift. Specifically, when the receiver and the transmitter move towards each other, the frequency of the EM waves received will increase; on the contrary, when they move away from each other, the frequency of the EM waves received will decrease.

The expression of the normalized Doppler frequency shift at a point on earth is given in \cite{DooplerLEO} as shown below.
\begin{equation}
\begin{split}
\frac{\triangle f}{f} = -\frac{1}{c}\frac{r_{E}r\sin(\psi(t)-\psi(t_{0}))}{\sqrt{r_{E}^{2}+r^{2}-2r_{E}r\cos(\psi(t)-\psi(t_{0}))}} \\
\times\frac{\cos(cos^{-1}(\frac{r_{E}}{r}\cos(\theta_{max}))-\theta_{max})\omega_{F}(t)}{\cos(cos^{-1}(\frac{r_{E}}{r}\cos(\theta_{max}))-\theta_{max})},
\end{split}
\end{equation}
where $\frac{\triangle f}{f}$ represents the normalized Doppler frequency shift, $c$ is the speed of light, $r_{E}$ denotes the radius of the earth, $r$ means the distance between the satellite and the earth's center, $\psi(t)-\psi(t_{0})$ is the angular distance between the position at time $t$ and the maximum elevation angle ($\theta_{max}$) of the satellite at time $t_{0}$, and $\omega_{F}(t)$ represents the angular velocity of the LEO satellite in the ECF frame.

\begin{table}[t]
 \centering
 \caption{Doppler frequency shift.} 
     \begin{tabular}{c|c|c|c}

    \hline
     \hline
       \makecell{ Frequency \\ (GHz)}  &  Max Doppler  &   Relative Doppler & \makecell{ Max Doppler \\ shift variation }\\
        \hline
         \hline
         
       2  &  +/- 48 kHz & 0.0024 \% & -544 Hz/s \\
    \hline
       20  &  +/- 480 kHz & 0.0024 \% & -5.44 kHz/s \\
    \hline
       30  &  +/- 720 kHz & 0.0024 \% & --8.16 kHz/s \\
    \hline
       3.5 (estimated)  &  +/- 84 kHz & 0.0024 \% & -952 Hz/s \\
    \hline
     \end{tabular}
 \end{table}

As shown in Table II \cite{3GPPLEO}, for the LEO satellites at 600 km altitude, the maximum Doppler shift can reach $\pm$48 kHz when the carrier frequency is 2 GHz. Furthermore, the maximum Doppler shift is proportional to the operating frequency. Therefore, it could be estimated that the Doppler shift of the LEO satellite at 600 km can reach $\pm$84 kHz at 3.5 GHz carrier frequency.

\section{Performance Analysis for Proposed Distributed Beamforming}
The superposition of electromagnetic waves is related to their polarization. Since the direction of polarization is assumed to depend on the direction of the satellite's movement, the distributed beamforming performance under different orbital directions is discussed as follows.

\subsection{Parallel orbits}
For the scenario of two satellites moving in the same direction, the composed $E$ field vectors and $H$ field vectors can be obtained through (15) and (16). Then, by calculating the Poynting vector, the corresponding received power enhancement can be obtained. 
\begin{equation} \label{eq1}\small
\begin{split}
\bm{S_{av}}&=\frac{1}{T}\int^{T}_{0}\sum_{m=1}^{2}(\bm{E_{my}}+\bm{E_{mz}})\times\sum_{m=1}^{2}\bm{H_{mx}}\rm{d} \textit{t}\\
&=\frac{1}{2Z_{0}}[(E_{1m}^{2}\cos\theta_{1}+E_{1m}E_{2m}\cos\theta_{1}\cos\Delta\varphi\\
&\quad\ +E_{1m}E_{2m}\cos\theta_{2}\cos\Delta\varphi+E_{2m}^{2}\cos\theta_{2})
 (-\bm{e_{z}})\\
&\quad\ +(E_{1m}^{2}\sin\theta_{1} + E_{1m}E_{2m}\sin\theta_{1}\cos\Delta\varphi\\
&\quad\ -E_{1m}E_{2m}\sin\theta_{2}\cos\Delta\varphi-E_{1m}^{2}\sin\theta_{2})(-\bm{e_{y}})],
\end{split}
\end{equation}
where $\bm{S_{av}}$ denotes the average Poynting vector over a period of time $T$. If we suppose that the two EM waves arriving at UE have the same energy as it can be dynamically adjusted by transmitting power, then there is
\begin{equation}
E_{1m} = E_{2m} = \sqrt{2}E_{0},
\end{equation}
where $E_{0}$ represents the effective value of the $E$ field. For convenience, (19) is considered a default condition in other scenarios. Besides, $\triangle \varphi$ means the phase difference between the two EM waves at the UE, whose value is
\begin{equation}
\triangle \varphi = \varphi_{d} - (\varphi_{1}-\varphi_{2}).
\end{equation}
If the phase difference is eliminated, namely $\triangle \varphi = 0$, (18) can be further simplified  as
\begin{equation} 
\bm{S_{av}} = \frac{2E_{0}^{2}}{Z_{0}}[(\cos\theta_{1}+\cos\theta_{2})(-\bm{e_{z}})+(\sin\theta_{1}-\sin\theta_{2})(-\bm{e_{y}})].
\end{equation}
From the above equation, it is clear that the received power at the target point depends on the angles $\theta_{1}$ and $\theta_{2}$. When $\theta_{1} = \theta_{2} = 0$, $\bm{S_{av}}$ could achieve its maximum value as follows.
\begin{equation}
\bm{S_{av}} = \frac{4E_{0}^{2}}{Z_{0}}(-\bm{e_{z}}).
\end{equation}

The condition of $\theta_{1} = \theta_{2} = 0$ is considered ideal and stringent. However, in practical scenarios, there are approaches to approximate this ideal situation. For instance, satellites can be positioned in different orbital shells at varying altitudes. Additionally, in regions close to orbital crossings and approaches, particularly in high latitudes, satellites can be spaced very closely together.

Compared to the power provided by one satellite, two satellites can improve the received power by four folds at maximum, or 6 dB in decibels. As the angle $\theta$ increases gradually, the superimposed received power decreases correspondingly. In addition, the radiated EM wave's polarisation direction impacts the composed EM fields. Its impact is similar to that of perpendicular orbits or intersecting orbits, which will be discussed in the following parts.

If the number of satellites participating in distributed beamforming is increased to $N$, then the maximum of the average Poynting vector could reach
\begin{equation}
\begin{split}
\bm{S_{av}}&=\frac{1}{T}\int^{T}_{0}\sum_{m=1}^{N}(\bm{E_{my}}+\bm{E_{mz}})\times\sum_{m=1}^{N}\bm{H_{mx}}\rm{d} \textit{t}\\
&=\frac{N^{2}E_{0}^{2}}{Z_{0}}(-\bm{e_{z}}).
\end{split}
\end{equation}
As shown above, the distributed beamforming technique can enhance the received power by up to $N^2$ folds. 

\subsection{Perpendicular orbits}
Given that the polarization direction of the beam is parallel to the moving direction of the satellite, if the orbits of the two collaborating satellites are perpendicular to each other, their EM field vectors should also be perpendicular to each other. Thus, when two satellites are moving in perpendicular orbits, their EM field components can be written as
\begin{equation}
\begin{cases}
    \bm{E_{1y}} = E_{1m} \cos\theta_{1} \sin(\omega t - k R_{1} + \varphi_{1}) (-\bm{e_{y}}) \\
    \bm{E_{1z}} = E_{1m} \sin\theta_{1} \sin(\omega t - k R_{1} + \varphi_{1}) (\bm{e_{z}}) \\
    \bm{E_{2x}} = E_{2m} \cos\theta_{2} \sin(\omega t - k R_{2} + \varphi_{2}) (-\bm{e_{x}}) \\
    \bm{E_{2z}} = E_{2m} \sin\theta_{2} \sin(\omega t - k R_{2} + \varphi_{2}) (\bm{e_{z}})
\end{cases}
\end{equation}

\begin{equation}
\begin{cases}
    \bm{H_{1x}} = \frac{E_{1m}}{Z_{0}} \sin(\omega t - k R_{1} + \varphi_{1}) (-\bm{e_{x}}) \\
    \bm{H_{2y}} = \frac{E_{2m}}{Z_{0}} \sin(\omega t - k R_{2} + \varphi_{2}) (\bm{e_{y}}).
\end{cases}
\end{equation}
Then, the average Poynting vector can be calculated as follows,
\begin{equation} \small
\begin{split}
\bm{S_{av}}&=\frac{1}{T}\int^{T}_{0}\sum_{m=1}^{2}(\bm{E_{mx}}+\bm{E_{my}}+\bm{E_{mz}}) \times\sum_{m=1}^{2}(\bm{H_{mx}}+\bm{H_{my}})\rm{d} \textit{t}\\
&=\frac{E_{0}^{2}}{Z_{0}}[(\cos\theta_{1}+\cos\theta_{2})(-\bm{e_{z}}) + (\sin\theta_{1} + \sin\theta_{2}\cos\Delta\varphi)(-\bm{e_{y}})\\
&\quad\ + (\sin\theta_{1}\cos\Delta\varphi + \sin\theta_{2})(-\bm{e_{x}})].
\end{split}
\end{equation}
The maximum average Poynting vector would be achieved when the two satellites are both right above the UE, namely $\theta_{1} = \theta_{2} = 0$, shown below.
\begin{equation}
\begin{split}
\bm{S_{av}}=\frac{2E_{0}^{2}}{Z_{0}}(-\bm{e_{z}}),
\end{split}
\end{equation}
which shows that the received power can only be increased to twice as much as when using a single satellite. Its corresponding decibel value is 3 dB, which is only half of that with parallel orbits. Besides, if there is a group of $N$ satellites working coherently, half of them are in parallel orbits, and the other half are in perpendicular orbits, then the result would be
\begin{equation}
\bm{S_{av}} = \frac{N^{2}E_{0}^{2}}{2Z_{0}}(-\bm{e_{z}}).
\end{equation}
From the result, it can be known that the maximum enhancement provided by $N$ satellites in perpendicular orbits is $\frac{N^{2}}{2}$. 

\subsection{Intersecting orbits}
In addition to parallel and perpendicular orbits, orbits can also intersect. Assuming that two satellites are moving in two orbits, one orbit is along with the $i$-axis, and the other orbit intersects it with an angle of $\xi$. Then the EM field components could be written as

\begin{equation}
\begin{cases}
    \bm{E_{1x}} = E_{1m} \cos\theta_{1} \cos\xi \sin(\omega t - k R_{1} + \varphi_{1}) (-\bm{e_{x}}) \\
    \bm{E_{1y}} = E_{1m} \cos\theta_{1} \sin\xi \sin(\omega t - k R_{1} + \varphi_{1}) (-\bm{e_{y}}) \\
    \bm{E_{1z}} = E_{1m} \sin\theta_{1} \sin(\omega t - k R_{1} + \varphi_{1}) (\bm{e_{z}}) \\
    \bm{E_{2x}} = E_{2m} \cos\theta_{2} \sin(\omega t - k R_{2} + \varphi_{2}) (-\bm{e_{x}}) \\
    \bm{E_{2z}} = E_{2m} \sin\theta_{2} \sin(\omega t - k R_{2} + \varphi_{2}) (\bm{e_{z}})
\end{cases}
\end{equation}

\begin{equation}
\begin{cases}
    \bm{H_{1x}} = \frac{E_{1m}}{Z_{0}} \sin\xi \sin(\omega t - k R_{1} + \varphi_{1}) (-\bm{e_{x}}) \\
    \bm{H_{1y}} = \frac{E_{1m}}{Z_{0}} \cos\xi \sin(\omega t - k R_{1} + \varphi_{1}) (\bm{e_{y}}) \\
    \bm{H_{2y}} = \frac{E_{2m}}{Z_{0}} \sin(\omega t - k R_{2} + \varphi_{2}) (\bm{e_{y}}).
\end{cases}
\end{equation}
The average Poynting vector can be calculated as
\begin{equation} \small
\begin{split}
\bm{S_{av}}&=\frac{1}{T}\int^{T}_{0}\sum_{m=1}^{2}(\bm{E_{mx}}+\bm{E_{my}}+\bm{E_{mz}}) \times\sum_{m=1}^{2}(\bm{H_{mx}}+\bm{H_{my}})\rm{d} \textit{t}\\
&=\frac{E_{0}^{2}}{Z_{0}}[(\cos\theta_{1}+\cos\theta_{1}\cos\xi\cos\Delta\varphi+\cos\theta_{2}\cos\xi\cos\Delta\varphi\\
&\quad\ +\cos\theta_{2})(-\bm{e_{z}}) + (\sin\theta_{1}\sin\xi + \sin\theta_{2}\sin\xi\cos\Delta\varphi)(-\bm{e_{y}})\\
&\quad\ + (\sin\theta_{1}\cos\xi + \sin\theta_{1}\cos\Delta\varphi + \sin\theta_{2}\cos\xi\cos\Delta\varphi\\
&\quad\ + \sin\theta_{2})(-\bm{e_{x}})].
\end{split}
\end{equation}
Similarly, its maximum can be obtained as below when $\Delta\varphi = 0$ and $\theta_{1} = \theta_{2} = 0$.
\begin{equation}
\bm{S_{av}} = \frac{(2+2\cos\xi)E_{0}^{2}}{Z_{0}}(-\bm{e_{z}}).
\end{equation}
If the number of satellites increases to $N$, and M satellites have an intersection angle of $\xi$, then the result becomes
\begin{equation}
\bm{S_{av}} = \frac{[N^{2}-MN(1-\cos\xi)]E_{0}^{2}}{Z_{0}}(-\bm{e_{z}}).
\end{equation}
The result shows that the enhancement is $N^{2}-MN(1-\cos\xi)$. Furthermore, the enhancement depends on the number of satellites moving in the intersecting orbits and their intersection angles.



\section{Numerical results}
\subsection{Comparison with MISO}
\begin{figure}
  \centering
  \includegraphics[width=3.3in]{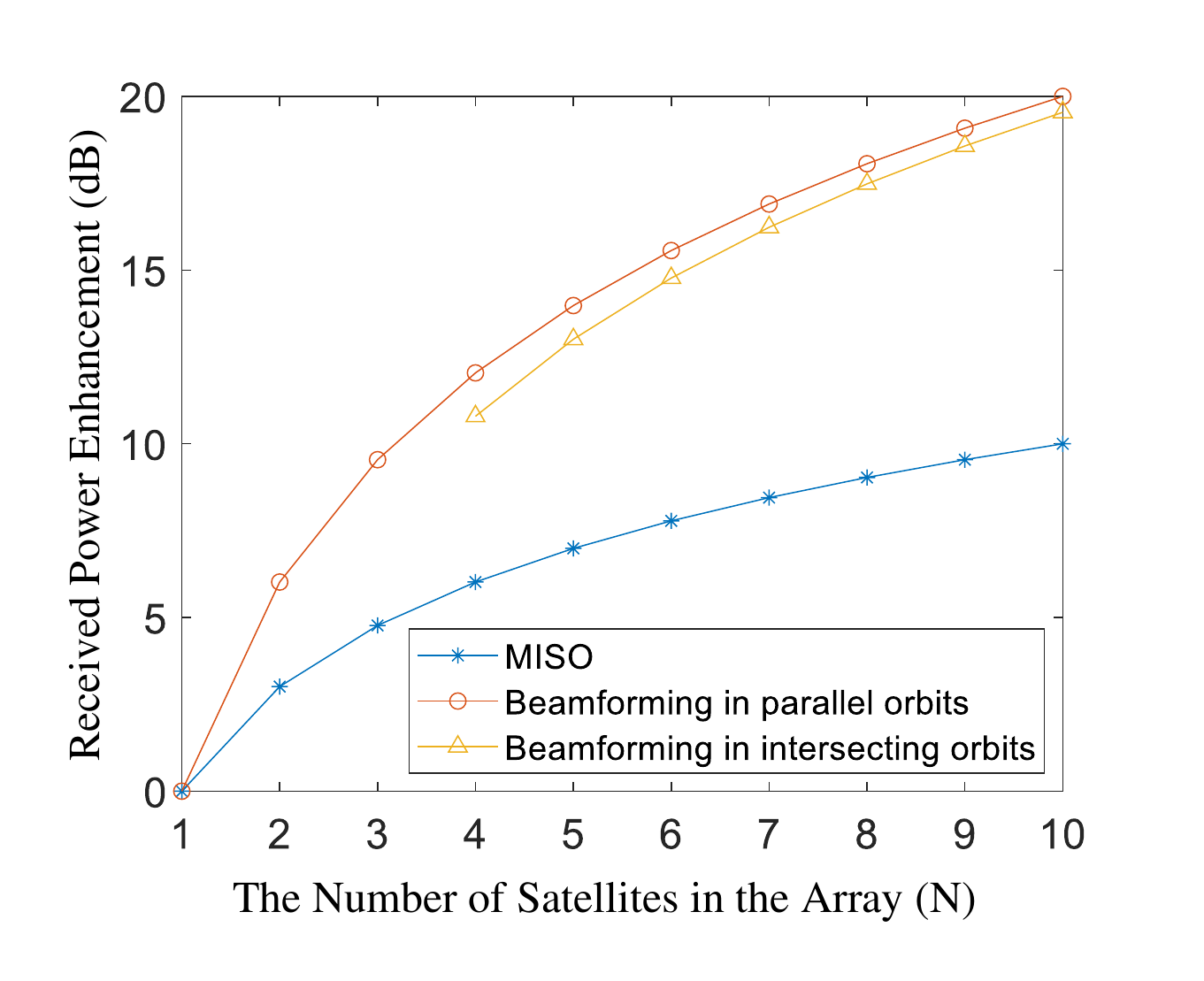}
  \caption{The comparison between distributed beamforming and MISO in the enhancement of received signal strength.}\label{BFvsMISO}
\end{figure}
Currently, the MIMO technique is not applicable to direct LEO satellite-to-smartphone communications due to the insufficient link budget. Assuming it is available in the future, its comparison with distributed beamforming is as follows. With the benefit of multi-antenna technology, a MIMO system can provide both diversity gain and multiplexing gain. Typically, its maximum diversity gain is proportional to the product of the number of transmitters and the number of receivers. Herein, we consider the smartphone with a single receiving antenna, so the MIMO model is simplified to a multiple-input single-output (MISO) model. Therefore, when there are $N$ transmitters, the maximum increment brought by MISO is $N$. As illustrated in Fig.~\ref{BFvsMISO}, the blue curve with stars represents the improvement by MISO, and the red curve with circles represents the increment by distributed beamforming in parallel orbits. As the number of satellites in the array increases, the received power enhancements from distributed beamforming and MISO increase, while the enhancement from the former in parallel orbits is twice as significant as that from the latter. In addition, the enhancement from distributed beamforming in intersecting orbits is also given, represented by the orange curve with triangles. It is assumed that two of the $N$ satellites are moving in the same intersecting orbit with an angle of 60 degrees, and the total number of satellites in the array starts at four. The figure shows that its enhancement can reach 10.8 dB when there are four satellites in the array in total. The gap between the two kinds of distributed beamforming scenarios gradually narrows as the total number of satellites in the array increases while the number of satellites in the intersecting orbit remains the same.

\subsection{Coverage patterns}
In practice, the coverage of satellite signals is a significant concern. Based on the analysis and calculation in the last section, various simulated coverage patterns from the perspective of different numbers and satellite deployments are presented in this section.

\subsubsection{Case 1 - A single satellite}

\begin{figure}
  \centering
  \includegraphics[width=3in]{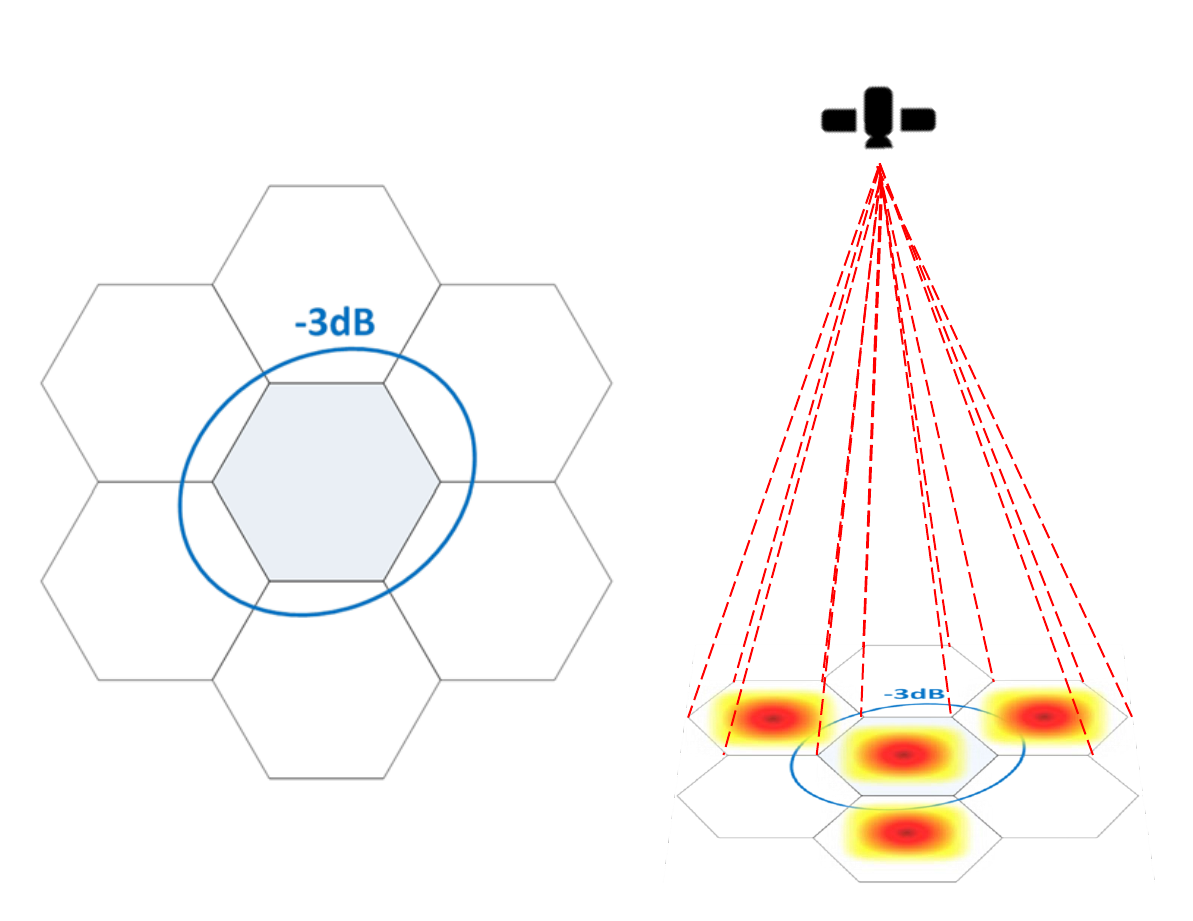}
  \caption{Intended beam coverage area with a single satellite.}\label{1Sat}
\end{figure}

Multi-beam radiation is commonly used by satellite communications nowadays. The intended beam coverage provided by a single satellite is shown in Fig.~\ref{1Sat}, where each beam is a cell inside the -3 dB contour. For instance, in the Starlink constellation, each beamwidth is 2.5 degrees and satellites run in orbits around 550 km. With these two parameters, it is accessible to obtain that the coverage of each beam is a circle with a radius of about 12 km. Moreover, the figure shows that the antenna gain achieves its maximum at the nadir and drops gradually as the beam slants away from the nadir. In other words, the received power is uniformly distributed within the cell. However, from the link budget discussed above, such a weak single cannot support higher transmission rates.

\subsubsection{Case 2 - Two satellites move in the same direction}
\begin{figure}
  \centering
  \includegraphics[width=3.3in]{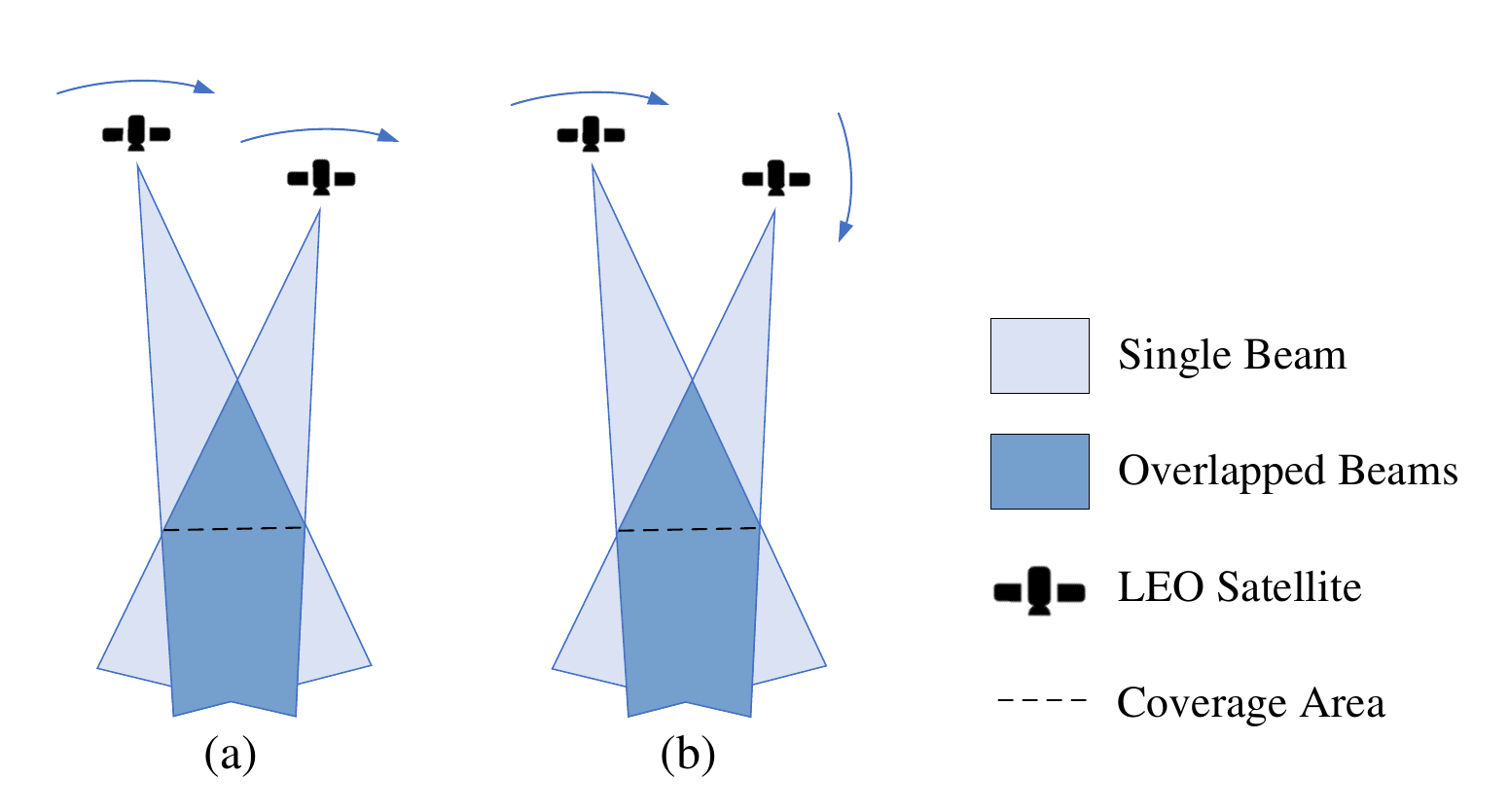}
  \caption{Two satellites work coherently: (a) Parallel orbits, (b) Vertical orbits.}\label{2Sat}
\end{figure}

In this case, two satellites running in parallel orbits work coherently. To compare with the cell beam coverage by a single satellite, distributed beamforming is implemented by selecting one beam from each satellite among its multiple beams. As depicted in Fig.~\ref{2Sat} (a), two satellites move in the same direction, and their beams overlap at certain distances. The dashed line represents the coverage area on the ground where the UE supposes to be located. The receiver could be deployed anywhere within the overlapped area marked by dark blue, but those applications are not included herein. In this case, the superposition between the two EM waves is similar to the model illustrated in Fig. 2. When two beams operating at the same frequency arrive at the receiver simultaneously, interference fringes will appear within the coverage area.
Furthermore, the fringes are perpendicular to the line between the two satellites. The width of the fringes depends on the incident angle of each beam. The smaller the angle between the beams, the wider the fringes are.
\begin{figure}
  \centering
  \includegraphics[width=3.4in]{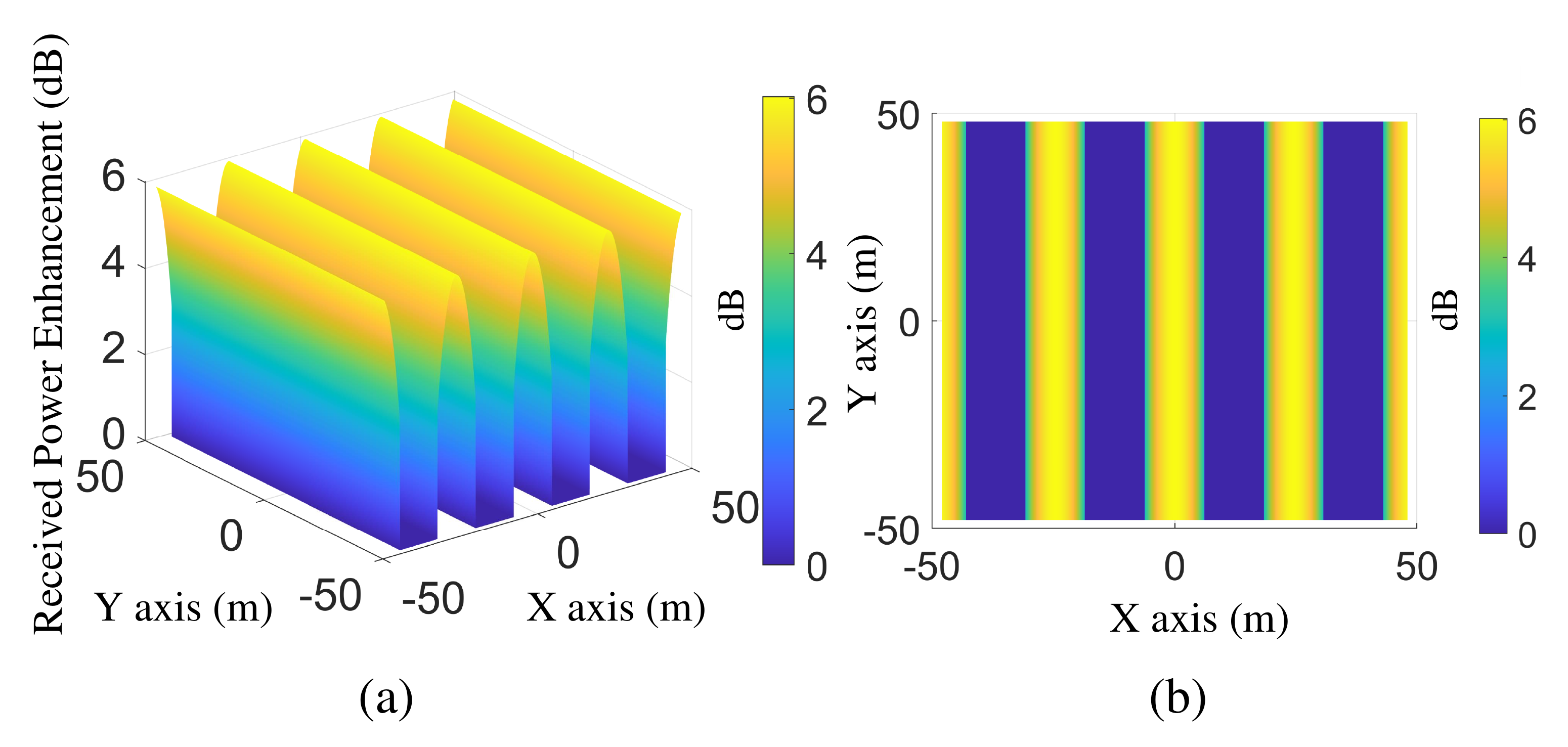}
  \caption{The coverage pattern of two satellites moving in parallel: (a) 3D view, (b) Top view.}\label{2parallel}
\end{figure}

As shown in Fig.~\ref{2parallel}, the coverage pattern is plotted in a square with a side length of 48 meters, taking UE as the center point. When the angle between the two incident beams is 0.2 degrees, the interference fringes are around 12 meters wide, and the lateral distance between satellites is about 1.92 km. The high-density satellites in LEO mega-constellations and multiple shells make the slight angle reasonable. In addition, the figure shows that a nearly 6 dB as maximum can be achieved by distributed beamforming.

\subsubsection{Case 3 - Two satellites run vertically}
\begin{figure}
  \centering
  \includegraphics[width=3.4in]{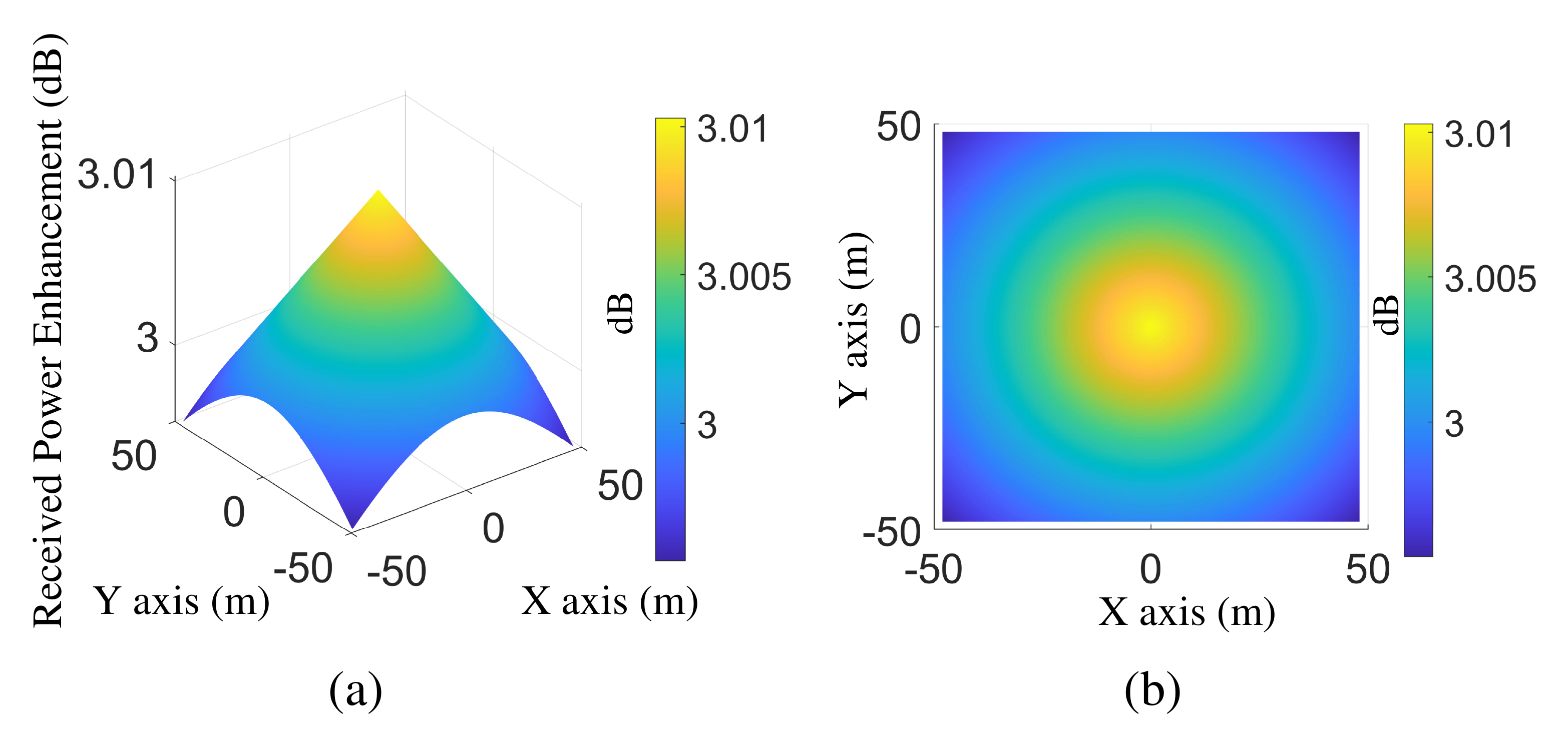}
  \caption{The coverage pattern of two satellites moving vertically: (a) 3D view, (b) Top view.}\label{2vertical}
\end{figure}

Two satellites turn round the earth, which may move upwards on one side of the earth and move downwards on the other side, causing some orbits to be perpendicular to others. This scenario is demonstrated by Fig.~\ref{2Sat} (b), where two satellites follow perpendicular trajectories. Instead of interference fringes appearing in the case of parallel orbits, the coverage pattern is similar to that of a single satellite. Fig.~\ref{2vertical} shows its 3D view on the left and its top view on the right. The pattern in its top view is a flat surface, and the received power enhancement gradually decreases as the position is away from the center point. Additionally, the EM field vectors of the two radiation waves are distributed in different planes due to their polarization and orbits. As a result, the received power enhancement can only reach a maximum of 3 dB. 

\subsubsection{Case 4 - Four satellites running parallel to each other}
\begin{figure}
  \centering
  \includegraphics[width=3.3in]{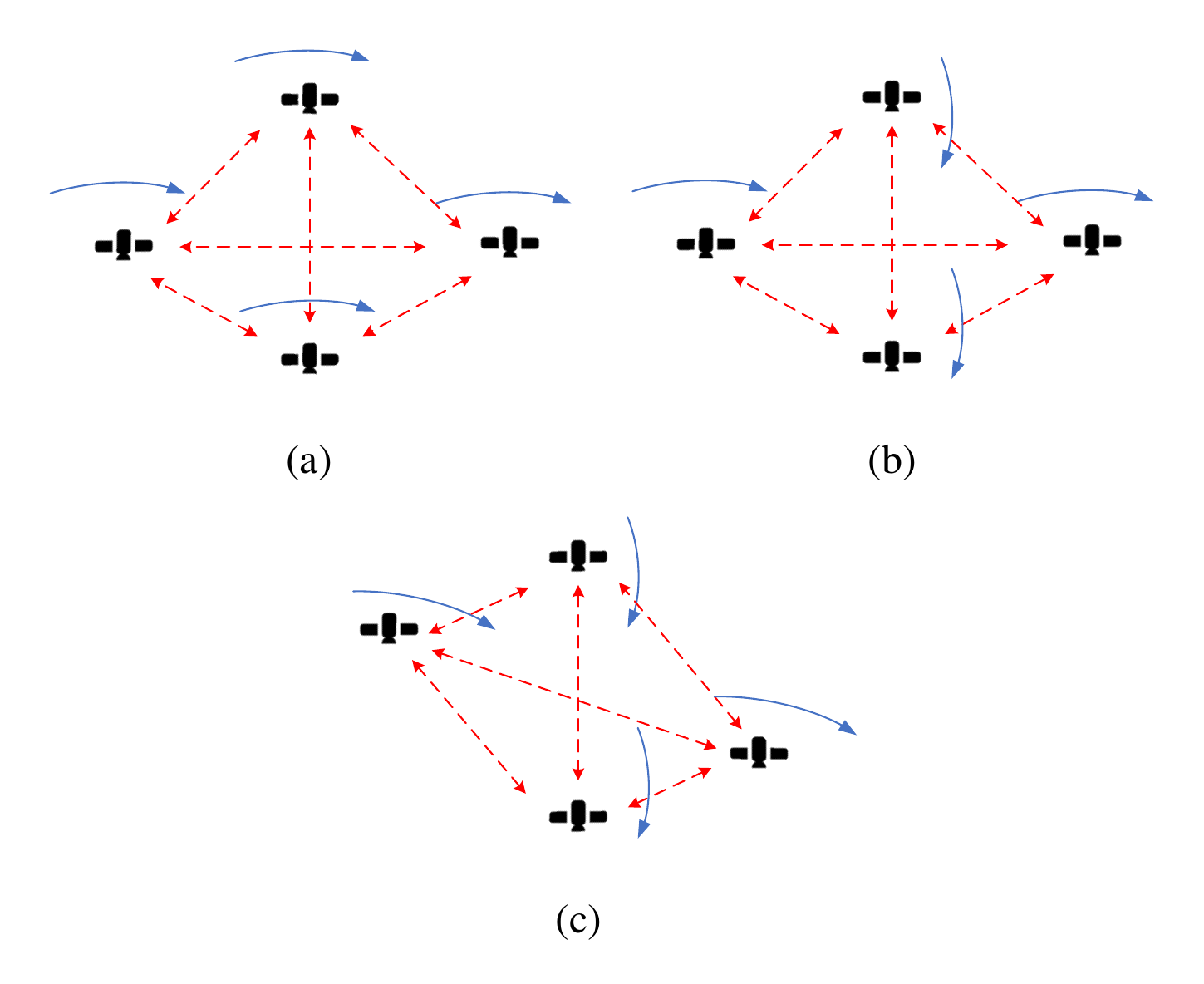}
  \caption{Four satellites work coherently: (a) Parallel orbits, (b) Vertical orbits, (c) Intersecting orbits.}\label{4Sat}
\end{figure}

\begin{figure}
  \centering
  \includegraphics[width=3.4in]{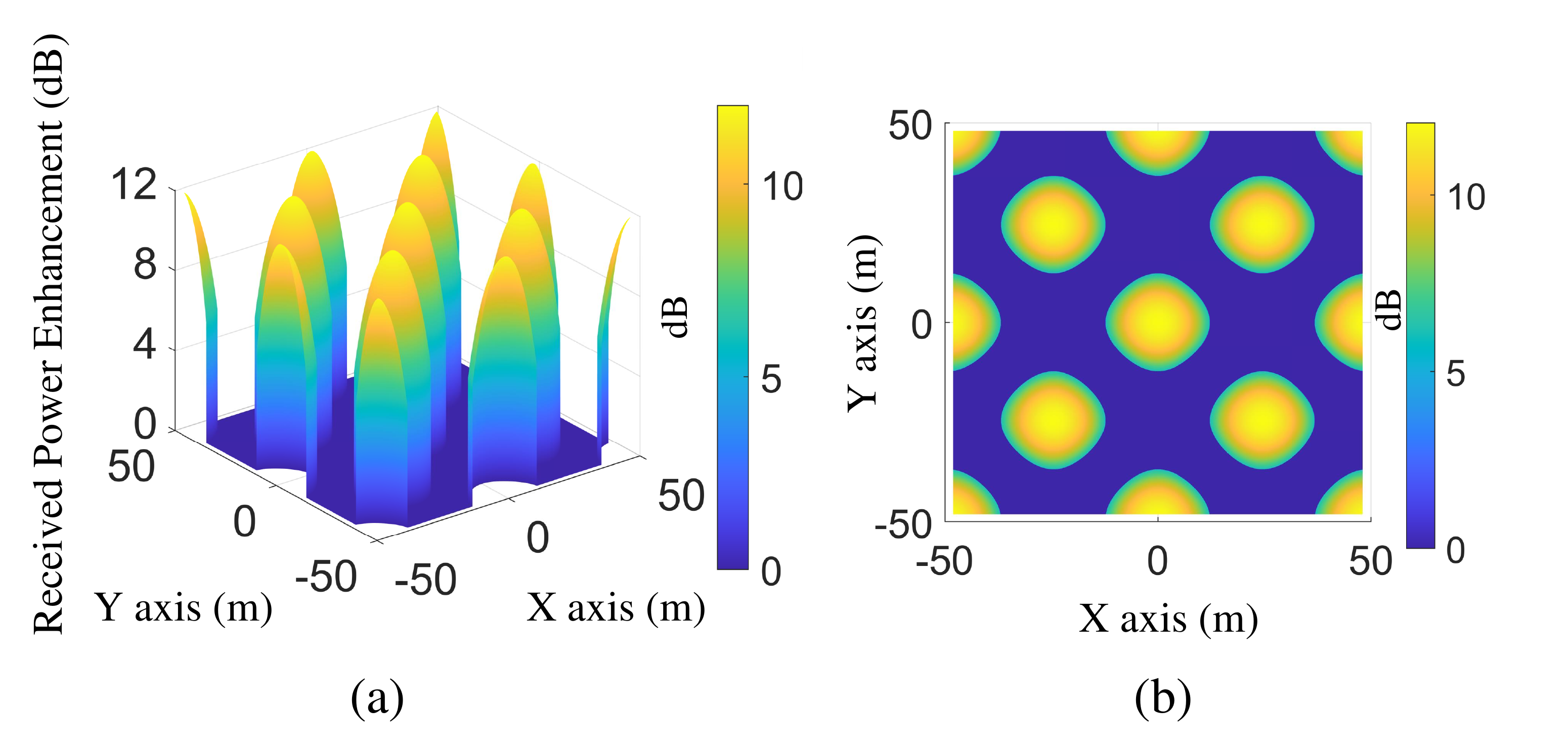}
  \caption{The coverage pattern of four satellites moving in parallel: (a) 3D view, (b) Top view.}\label{4parallel}
\end{figure}

This case describes a four-element array with parallel orbits as shown in  Fig.~\ref{4Sat} (a). For simplicity, we assume that four satellites are symmetrically located about the UE in a top view. Such a geometry yields a spot-beam pattern that is axially symmetric and centrally symmetric. From the picture, we can also realize that the maximum received power enhancement achieves 12 dB. As known to all, diversity gain is proportional to the number of radiation elements, corresponding to 4 folds or 6 dB in this case. For comparison, the power enhancement in Fig.~\ref{4parallel} is cut off at 6 dB. If the spot is considered a circle, its radius is around 12 m, and its area is around 452 m$^2$.

\subsubsection{Case 5 - Four satellites running vertically to each other}
\begin{figure}
  \centering
  \includegraphics[width=3.4in]{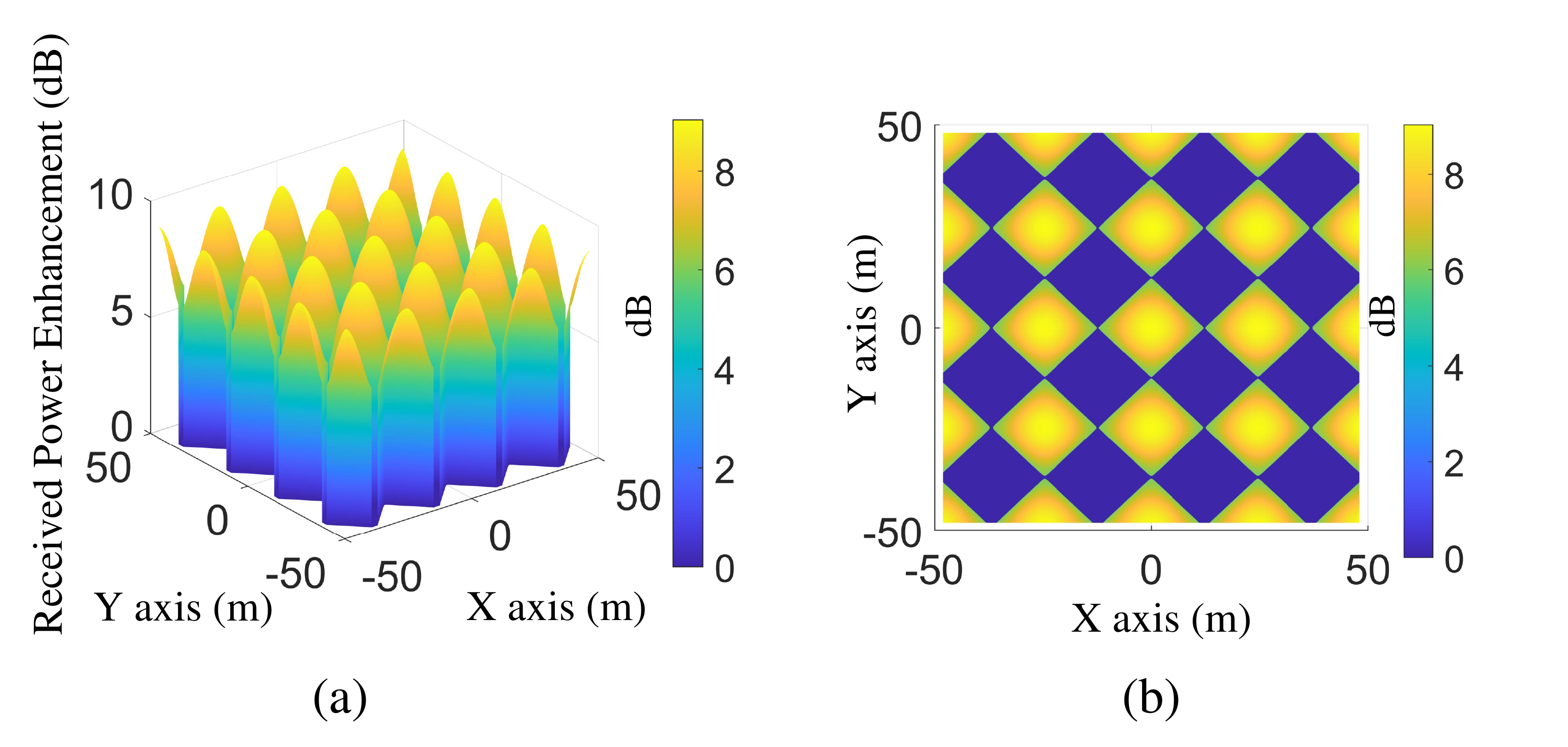}
  \caption{The coverage pattern of four satellites moving vertically: (a) 3D view, (b) Top view.}\label{4vertical}
\end{figure}

In this case, the geometry of each satellite's distributed array and parameters is similar to those in Case 4, while the difference is that orbits are perpendicular to each other, as shown in Fig.~\ref{4Sat} (b). Its coverage pattern in the 3D view and the top view is shown in Fig.~\ref{4vertical}. Similarly, the data is cut off at 6 dB. It can be observed that the maximum received power enhancement reaches 9 dB and the shape of the pattern is more like squares. From another perspective, the pattern seems to be formed by rotating the pattern of Case 2 and then superimposing it with itself. The diagonal length of the square is around 24 m, and the area is about 288 m$^2$. Compared with Case 4, the maximum received power enhancement, in this case, is 3dB less. The fundamental reason is that the direction of polarization is vertical, which causes the amplitude of the synthesized EM field vectors to be $\sqrt{2}$ times lower than the maximum.

\subsubsection{Case 6 - Four satellites move in two intersecting orbits}
\begin{figure}
  \centering
  \includegraphics[width=3.4in]{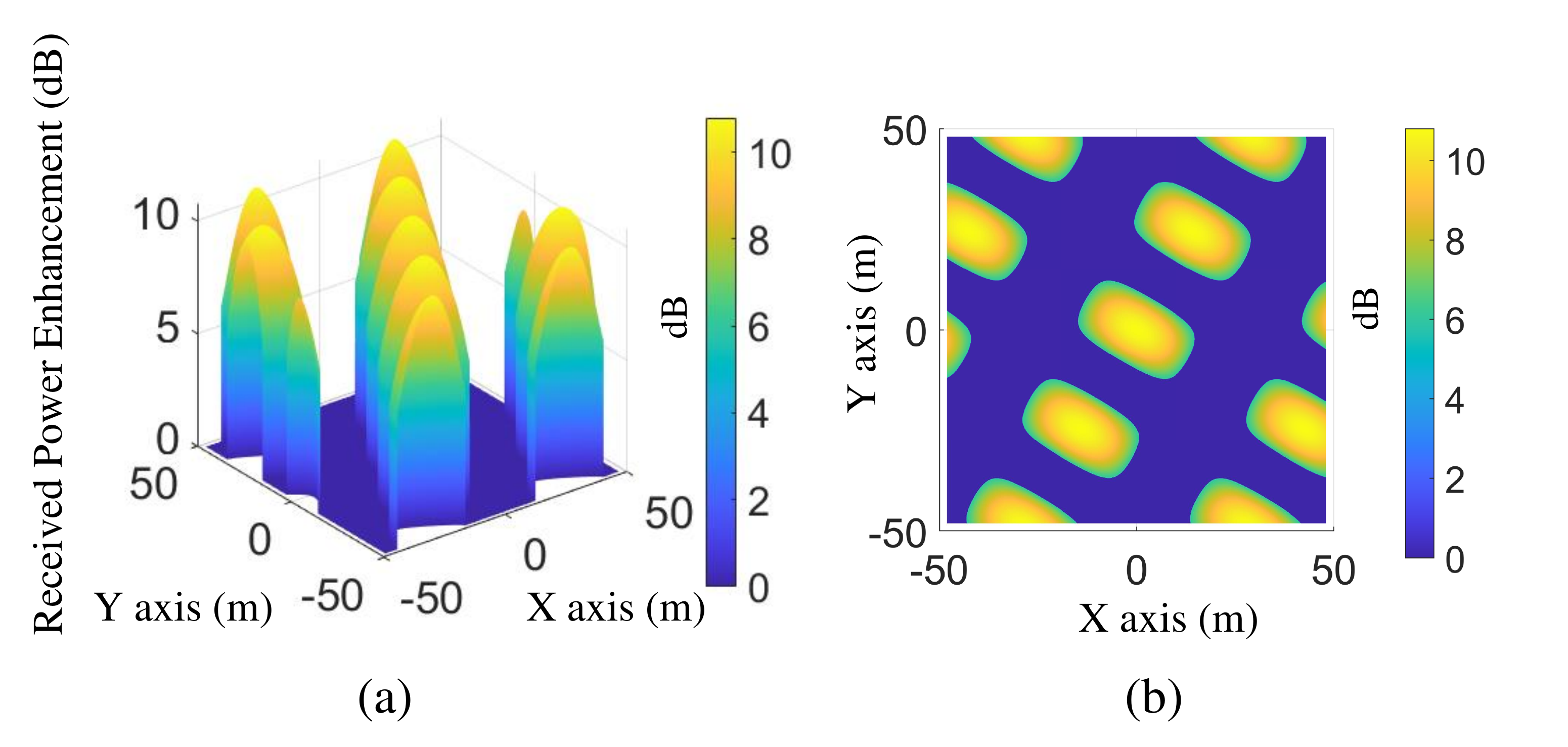}
  \caption{The coverage pattern of four satellites moving in intersecting orbits: (a) 3D view, (b) Top view.}\label{4intersect}
\end{figure}
A more general case is illustrated by Fig.~\ref{4Sat} (c), where four satellites move in two intersecting orbits. When the intersection angle is set to 60 degrees, the simulation results of the coverage pattern in the 3D view and the top view are presented in Fig.~\ref{4intersect}. It shows that the maximum received power enhancement is around 10.8 dB. The top view indicates that the shape of the pattern looks like rounded rectangles, and there is a certain angle between the pattern and the axis due to the orbits. Because the intersection angle is between 0 and 90 degrees, the coverage area and the enhancement of received power is smaller than those in Case 4 and more significant than those in Case 5.

\subsection{Doppler effect on distributed beamforming}
Typically, the Doppler effect is researched on a single satellite. But considering the proposed satellite collaborative model, the frequency shift between satellites also needs to be studied. Herein, two satellites are assumed to be working at different frequencies. Each satellite transmits one beam, and its EM waves work constructively in the target area. According to the analysis in Section II, it is known that the beamforming performance relies on the extent to which the EM waves interfere.

\begin{figure}
  \centering
  \includegraphics[width=3.1in]{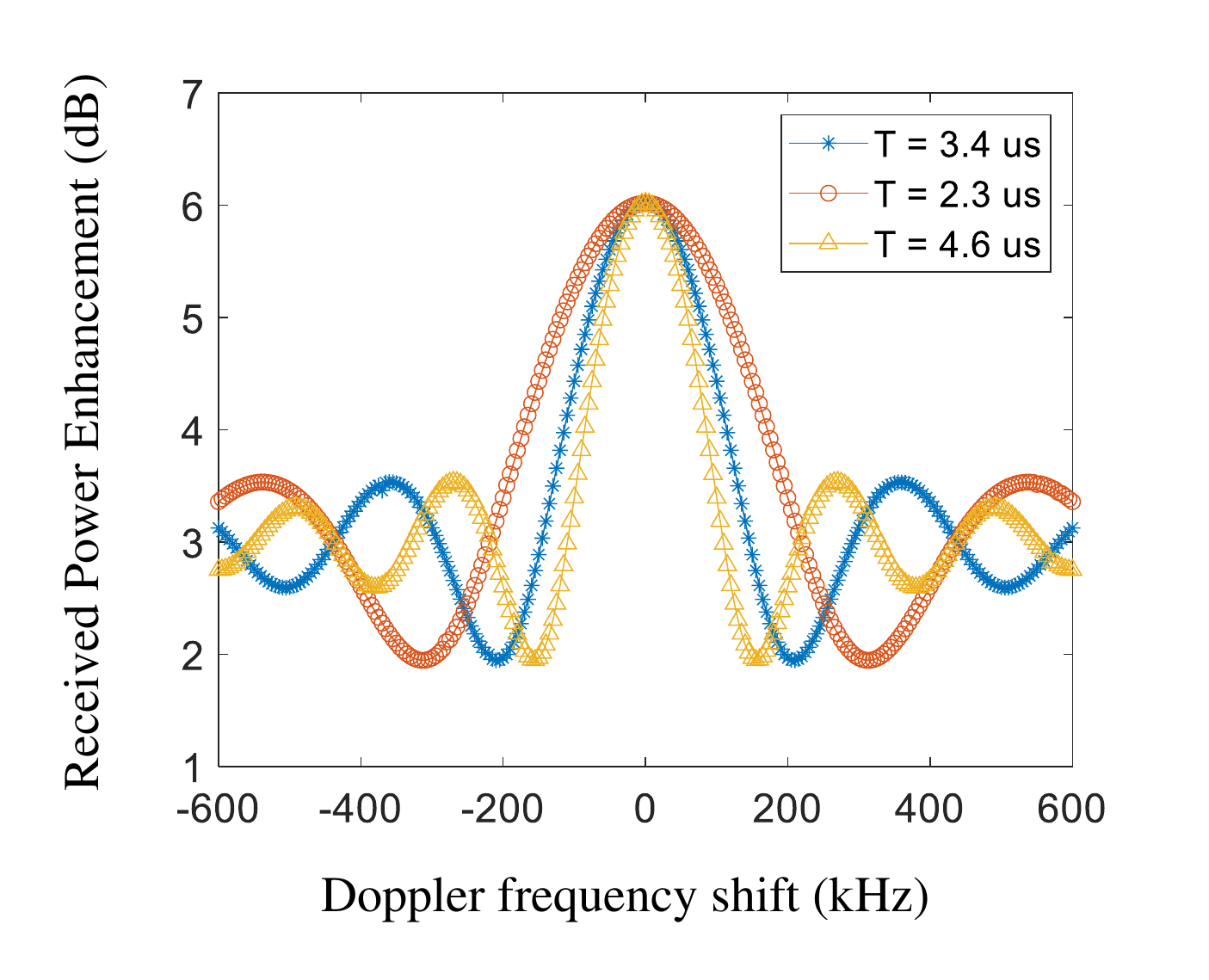}
  \caption{Doppler effect on distributed beamforming performance.}\label{Dop}
\end{figure}

In practice, each beam has its Doppler frequency shift. For convenience, the frequency of the EM waves from Sat1 is used as a reference, and all the frequency shift is concentrated on the beam transmitted by Sat2. Specifically, Sat1 is set to operate at 3.5 GHz, and Sat2 is set to operate at 3.5$\pm \triangle f $ GHz, where $\triangle f$ represents the frequency shift. As shown in Fig.~\ref{Dop}, the horizontal axis represents the frequency shift, and the vertical axis represents the enhancement of received power in dB. The maximum of 6 dB can be achieved when $\triangle f = 0$, which means no frequency or Doppler shift is completely compensated for. This value is consistent with the results of the previous analysis. The distributed beamforming performance deteriorates with the increase of frequency shift, and the performance curve oscillates along the horizontal axis and gradually flattens out. 

From  (18), it is known that the calculation result also depends on the period $T$. $T$ is first set to 3.4 $\upmu$s which corresponds to 12000 wave periods. If each period can represent a bit, 12000 periods can represent 1500 bytes, approximately the maximum transmission unit (MTU) in the link layer. As can be observed from the blue curve with stars in the figure, to attain an enhancement higher than 3 dB, the Doppler shift needs to be less than 146 kHz. Moreover, as $T$ decreases, the curve flattens out, as presented by the red curve with circles.
Conversely, as $T$ increases, the curve oscillation becomes increasingly violent, as shown by the orange line with triangles. For example, two sinusoidal waves with different frequencies propagate in the same direction and have the same phase at the starting point. In the beginning, two waves superimpose together, and the result of the integral in one period is greater than that of a single beam. At some point, however, the integral of the superimposed beam is equal to that of a single beam. And even worse, when two waves are out of phase, the integral may even be zero. 

\subsection{The effect of time misalignment}
\begin{figure}
  \centering
  \includegraphics[width=3.4in]{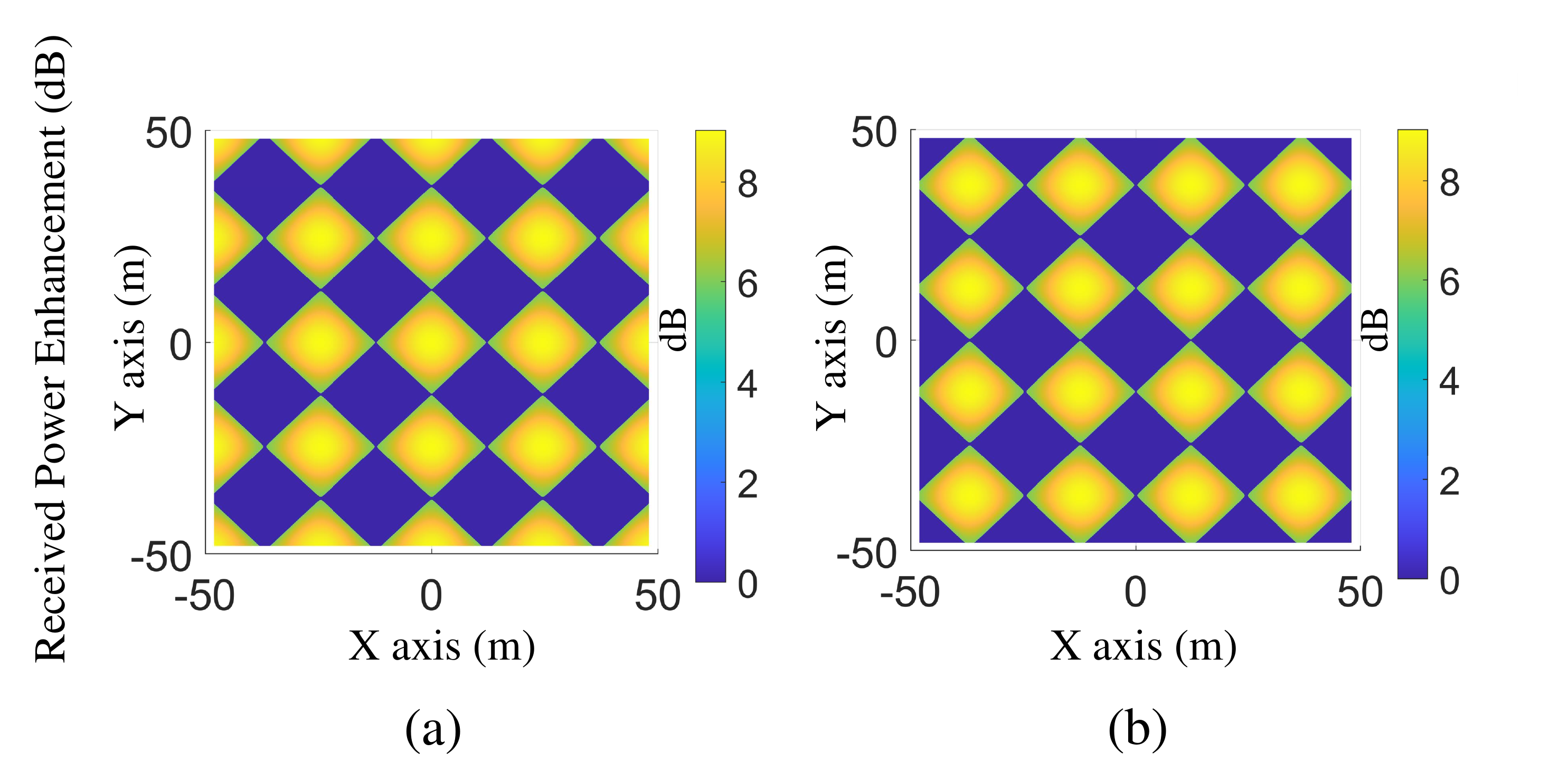}
  \caption{The coverage pattern generated by four satellites moving vertically: (a) EM waves arrive simultaneously, (b) With half a period arrival time difference.}\label{TimeMis}
\end{figure}

Time alignment is one of the most fundamental conditions to realize distributed beamforming because it relies on interference between EM waves. If there is time misalignment between the receiving EM waves which means some waves arrive first and others arrive later, then the EM wave interference at the receiver would not be as expected. But the interference still exists if frequency synchronization is satisfied since we consider signal coverage rather than a single receiver. The time synchronization problem could be attributed to the phase synchronization problem. Assuming that two EM waves have the same initial phase when they arrive at the receiver, but the arrival time is half a period apart, they would be out of phase when they interfere. To show the comparison of the coverage pattern caused by the time misalignment, the coverage patterns in the top view generated by four satellites moving vertically are presented in Fig.~\ref{TimeMis}. (a) illustrates the pattern where four beams arrive simultaneously, while (b) depicts that there is half a period arrival time difference between each. By contrast, the pattern looks like it is shifted, although the maximum enhancement is about the same. Thus, the composed beam may not cover the aimed area. In addition, LEO satellites move at a high velocity resulting in a short coordinating time, and the time misalignment would cause that information cannot be delivered promptly.

\section{Conclusion}
The paper proposed a novel distributed beamforming technique based on the LEO mega-constellation to enhance the received signal strength and achieve direct transmission between LEO and smartphones. We then theoretically analyzed the superposition of the EM waves in detail. When there are $N$ satellites working coherently, the received power can be enhanced up to $N^2$ folds or $20log(N)$ dB through distributed beamforming. In contrast, only $N$-fold enhancement could be obtained using the MISO technique even though it could be available. In addition, a variety of coverage patterns were provided by simulation, considering the different numbers of satellites and various deployments. Time alignment, frequency synchronization, and phase calibration are necessary to achieve distributed beamforming. Thus we also discussed the impact of these factors on the distributed beamforming performance, especially in terms of the Doppler effect and time misalignment.

\ifCLASSOPTIONcaptionsoff
  \newpage
\fi

\bibliographystyle{IEEEtran}
\bibliography{IEEEabrv,Bibliography}

\vfill


\end{document}